\documentclass[9pt,twoside]{opticajnl}
\journal{opticajournal} 

\setboolean{shortarticle}{false}

\usepackage{lineno}
\usepackage{multicol}
\usepackage{lipsum}
\usepackage{siunitx}
\usepackage[hypcap=true]{caption}

\makeatletter
\newcommand{\labeltext}[2]{%
  \@bsphack
  \MakeLinkTarget*{#1}%
  \def\@currentlabel{#1}{\label{#2}}%
  \@esphack
}
\makeatother

\newenvironment{Figure}
  {\par\medskip\noindent\minipage{\linewidth}}
  {\endminipage\par\medskip}

\newcommand{\Rb}[0]{${}^{85}$Rb}
\newcommand{\Cs}[0]{${}^{133}$Cs}

\DeclareSIUnit{\belmilliwatt}{Bm}
\DeclareSIUnit{\dBm}{\deci\belmilliwatt}
\DeclareSIUnit\bar{bar}

\title{Simultaneous optical and multi-band terahertz imaging using an atomic quantum sensor.}
\author[1]{Andrew Rae MacKellar*}
\author[1]{C. Stuart Adams}
\author[1]{Kevin J. Weatherill}

\affil[1]{Quantum Light and Matter Group, Department of Physics, Durham University, South Road, Durham DH1~3LE, United Kingdom}

\affil[*]{andrew.r.mackellar@durham.ac.uk}

\begin{abstract}
We demonstrate simultaneous imaging at 0.549\,THz and 1.012\,THz with an optical overlay using a two-species atomic-vapour-based technique. The atomic vapour, comprising laser-pumped rubidium and caesium atoms contained within the same cell, is used to convert two narrowband terahertz signals to optical frequencies which can then be detected using standard CMOS sensors. We use the system to image and perform spectral analysis of material samples. As atomic vapour is optically transparent, by using optically-transparent terahertz lenses, we can achieve simultaneous optical imaging, allowing for potential integration of terahertz sensitivity into a range of optical imaging devices.  
\end{abstract}

\setboolean{displaycopyright}{false} 
\begin{document}
\maketitle
\begin{multicols}{2}
\section{Introduction}

Hyperspectral and multispectral imaging techniques enable simultaneous acquisition of spatial and spectral information, allowing for the identification and classification of materials based on their spectral signatures. These techniques are widely employed in applications such as food quality assessment~\cite{gowen2007,Udayanga2024,Lee2023}, environmental monitoring~\cite{Kvaterniuk2019,Arnold2010}, security inspection~\cite{Vetrekar2024}, and biomedical imaging~\cite{lu2014,Goessinger2024,Feiyan2023,Ozkan2023}. Conventional multispectral systems typically operate in the visible and near-infrared (VIS–NIR) spectral regions, where material discrimination is achieved by leveraging reflectance differences across discrete wavelength bands~\cite{VanBeek2013,Stow2019,Burgos-Fernandez2022}.

Terahertz (THz) imaging, covering the spectral range from approximately 0.3 to 3\,\unit{\THz}, offers complementary capabilities to VIS–NIR modalities. Owing to the non-ionizing nature of THz radiation and its ability to penetrate many non-metallic, non-polar materials~\cite{Wu2024,Yehao2024}, THz imaging enables non-destructive probing of subsurface structures~\cite{federici2005,Medjadba2025}. Furthermore, numerous materials exhibit distinct spectral features in the THz range due to molecular vibrational and rotational resonances, providing a basis for chemical identification~\cite{tonouchi2007,Bratu2024,Singh2022}. These properties have led to increasing interest in THz multispectral and hyperspectral imaging for applications in agriculture~\cite{Mathialagan2025,Wang2018}, cultural heritage diagnostics~\cite{pickwell2006,Bertling2024,Vazquez2024,Groves2009}, medical imaging (e.g., cancer margin detection)~\cite{Yan2022,Taylor2011,Sung2012}, and the identification of concealed substances and illicit materials~\cite{chan2007,Kawase2003}.

Several imaging modalities have been developed for the THz domain, each optimized for specific application requirements. Time-domain spectroscopy (THz-TDS) is commonly employed when high spectral resolution and broadband coverage are needed~\cite{jepsen2011,Eliet2022,Nikolaev2020}. However, TDS-based imaging typically requires raster scanning, resulting in slow acquisition times for two-dimensional images.

In contrast, focal plane array (FPA) detectors, such as microbolometer and field-effect transistor (FET) arrays, enable rapid acquisition of two-dimensional images in a single exposure~\cite{kojima2017}. While advantageous for high-throughput imaging, these systems generally lack intrinsic spectral resolution and may exhibit limitations in sensitivity and temporal response.

Hybrid imaging architectures have emerged to address the trade-offs between spectral fidelity and acquisition speed. Such systems combine FPAs with tuneable THz sources or discrete-frequency THz emitters to enable multispectral imaging with improved temporal performance~\cite{chan2014}. These approaches offer a promising route toward practical, real-time THz imaging systems capable of balancing spectral and spatial resolution with acquisition throughput. However, the speed and sensitivity of the THz camera remains the limitation in these systems. 

Recently, new approaches to terahertz sensing have emerged using atom-based quantum sensors~\cite{wade2017}. Atomic THz sensors generally operate by mapping incoming THz signals on to infrared or optical fields. Previous work has demonstrated high-speed full-field imaging at kilohertz frame rates~\cite{downes2020,downes2023,Li_2025} and has enabled a range of optical techniques to be transferred to the THz range, such as super-resolution imaging~\cite{Fleming25} and fast readout of orbital angular momentum beams~\cite{Downes22}.

Because these atom-based sensors use narrowband atomic transitions that are well described by theory~\cite{sibalic2017}, they are intrinsically calibrated and can provide electric-field measurements that are linked to SI-units~\cite{chen2022, krokosz2025electricfieldmetrologyterahertzfrequency}. However, due to the discrete resonant frequency of the atomic transition they are limited to operation at particular discrete frequencies and, although rapid\end{multicols}
\begin{figure*}[t]
\centering
\includegraphics[width=\linewidth]{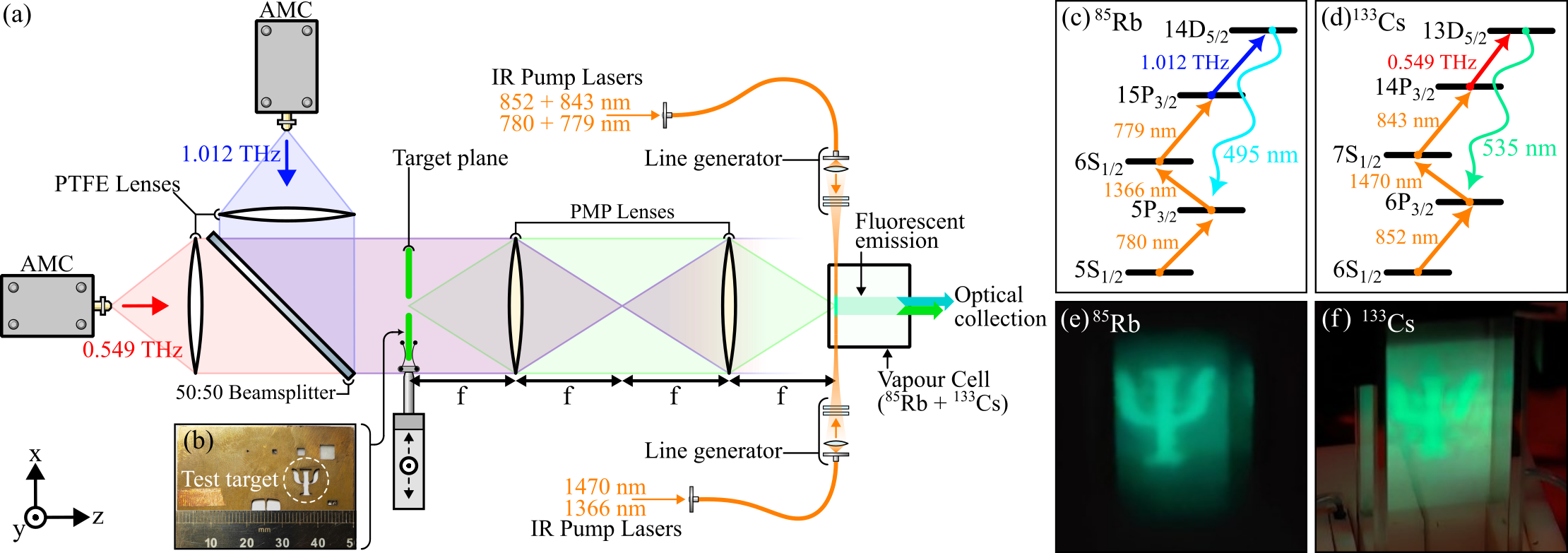}
\caption{Experimental setup of the two-species imager (a). Two fields at 0.549\,\unit{\THz} (red) and 1.012\,\unit{\THz} (blue) are collimated with f=75~mm lenses, and then overlapped by a high resistivity float zone (HRFZ) Si beamsplitter such that the beams are co-linear. This combined bi-chromatic field interacts with a target object, and is then re-imaged by a polymethylpentene (PMP) lens relay such that light from the target plane is refocussed into the science cell at the position of a light sheet (orange). This light sheet excites the ${}^{85}$Rb \& ${}^{133}$Cs atoms to states with transitions that are resonant with the 1.012\,\unit{\THz} \& 0.549\,\unit{\THz} fields. In the presence of these fields, the atoms fluorescence at 495\,\unit{\nm} (c) and 535\,\unit{\nm} (d), respectively. Real-colour images of the optical fluorescence when imaging a `$\Psi$' pattern from a metal test card (b) are shown in ${}^{85}$Rb (e) and ${}^{133}$Cs (f).}
\label{fig:expt_setup}
\end{figure*}
\labeltext{\ref{fig:expt_setup}\,a}{fig_expt_equipment}
\labeltext{\ref{fig:expt_setup}\,c}{fig_expt_decay_Rb}
\labeltext{\ref{fig:expt_setup}\,d}{fig_expt_decay_Cs}
\labeltext{\ref{fig:expt_setup}\,c,d}{fig_expt_decay_both}
\labeltext{\ref{fig:expt_setup}\,b}{fig_expt_testcard}
\labeltext{\ref{fig:expt_setup}\,e}{fig_expt_realcol_Rb}
\labeltext{\ref{fig:expt_setup}\,f}{fig_expt_realcol_Cs}
\labeltext{\ref{fig:expt_setup}\,e,f}{fig_expt_realcol_both}
\begin{multicols}{2}
\noindent switching between frequencies is possible~\cite{downes25}, up until now, only a single frequency at a time has been demonstrated. Here, we expand upon the operation of an atomic-vapour-based THz sensors: firstly; by using vapour comprised of two different atomic species, we can operate at two different THz frequencies simultaneously. Secondly, since the atomic vapour is transparent to optical frequencies, by using optically transparent THz lenses, we can also achieve a simultaneous optical overlay. We use the system to perform proof-of-principle materials analysis by differentiating between samples of different sugars.  

\section{Experimental setup}

\labeltext{\ref{fig:expt_setup}\,a}{fig_expt_a}
\labeltext{\ref{fig:expt_setup}\,d}{fig:expt_d}
\labeltext{\ref{fig:expt_setup}\,b}{fig:expt_setup_testcard}
The experimental setup is shown in Figure~\ref{fig_expt_a}. Terahertz fields are emitted by Virginia Diodes Amplifier Multiplier Chains (VDI AMC331 \& AMC731), supplied with radio frequencies generated by a Windfreak Technologies SynthHD PRO (v2) dual channel RF signal generator. One RF channel supplies $10,159.54\,\si{\MHz}$ at $10\,\si{\dBm}$ to the AMC331 ($\times$54 frequency multiplication) to output approximately $15\,\unit{uW}$ at 0.549\,\si{\tera\hertz}, with the field emission from a VDI WR1.5H (25) horn. The second channel supplies $10\,\si{\dBm}$ at $14,059.13\,\si{\MHz}$ to the AMC731 ($\times72$ frequency multiplication) to generate approximately $190\,\unit{uW}$ at 1.012\,\si{\tera\hertz}, emitted via a VDI WR1.0DH horn. Both fields have a 10${}^{\circ}$ full-width 3\,dB divergence angle, and are vertically polarised. The terahertz fields are collimated with f=75\,mm PTFE (Polytetrafluoroethylene) lenses, then overlapped with a HRFZ (High-Resistivity Float-Zone) silicon beam-splitter with a transmission:reflection ratio of approximately 55:45, generating a combined field that is projected along a main imaging axis (Figure~\ref{fig:expt_setup}, Z-axis). 

This combined, bi-chromatic field illuminates a target object (Figure~\ref{fig_expt_testcard}), located at one focus of a 4-f imaging relay comprised of two f=65\,mm polymethylpentene (PMP) lenses (Thorlabs 	
AL265, polished to enable optical imaging), which refocus the THz fields into a vapour cell filled with atomic rubidium (natural abundance) and atomic caesium-133, configured to a magnification ratio of approximately 1:1, resulting in an effective field of view of 1\,\unit{\cm\squared}. This target object is mounted on a motorised (Figure~\ref{fig:expt_setup}, X-Y) translation stage to expand the accessible field of view beyond the limits of the vapour cell by rastering the target position within the imaging field plane. The vapour cell is constructed with a quartz glass Hellma fluorescence cuvette (Part no. 101-10-40), with an internal path length of 10$\times10$\,$\si{\mm}$, filled and sealed under vacuum ($\approx{10}^{-7}\,\si{\m\bar}$). 
To generate approximately equal partial vapour pressures of each atomic vapour at typical experimental cell temperatures ($\approx60\,\unit{\celsius}$), the cell has been filled with a ratio of approximately 2:1 Rb:Cs metal (by mass), accordingly with Raoult's law~\cite{Foulkes-PhysChem2013}.

To prepare the atoms in the vapour in THz-sensitive states we use a total of six excitation lasers.
The excitation laser beams within the vapour cell are shaped by a pair of fibre-coupled light-sheet generators, comprised of a bare single-mode fibre-end followed by three cylindrical lenses. The first lens sets the working distance of the generator and the horizontal thickness of the light sheet, while the remaining two lenses first expand and then collimate the beam's vertical aspect. These lenses are configured such that they create a thin sheet of light within the cell of approximately 15\,\unit{\mm} height ($y$) and approximately 80\,\unit{\um} thickness ($z$), while the width of 10\,\unit{\mm} ($x$) is determined by the internal dimensions of the cuboidal science cell.

\labeltext{\ref{fig:expt_setup}\,c,d}{fig_expt_pathways}
The line generators are used to deliver light at wavelengths (with typical post-fibre powers) of; 852\,\unit{\nm} (16\,\unit{\mW}) , 1470\,\unit{\nm} (27\,\unit{\mW}), 843\,\unit{\nm} (130\,\unit{\mW}), 780\,\unit{\nm} (15\,\unit{\mW}), 1366\,\unit{\nm} (14\,\unit{\mW}), and 779\,\unit{\nm} (77\,\unit{\mW}). These wavelengths correspond to transitions within the atomic vapours; the \Cs{} is pumped along the excitation pathway ${6\textrm{S}}_{1/2} \rightarrow {6\textrm{P}}_{3/2} \rightarrow {7\textrm{S}}_{1/2} \rightarrow {14\textrm{P}}_{3/2}$, and similarly the \Rb{} is pumped via ${5\textrm{S}}_{1/2} \rightarrow {5\textrm{P}}_{3/2} \rightarrow {6\textrm{S}}_{1/2} \rightarrow {15\textrm{P}}_{3/2}$ (see Figure~\ref{fig_expt_pathways}). 
When the atoms within the cell move into the light sheet, they are thereby optically pumped into excited 'Rydberg' states that have resonant transitions to nearby states with frequencies  0.549\,\unit{\THz} (in \Cs{}) and 1.012\,\unit{\THz} (in \Rb{}) to match the incoming THz fields. When these atoms are simultaneously exposed to both the IR pump lasers and THz fields,  
\labeltext{\ref{fig:expt_setup}\,e,f}{expt_setup_ef} 
\labeltext{\ref{fig:expt_spectrum}\,a,b}{fig:expt_spectrum_ab}
\begin{Figure}
\centering
\includegraphics[width=\linewidth]{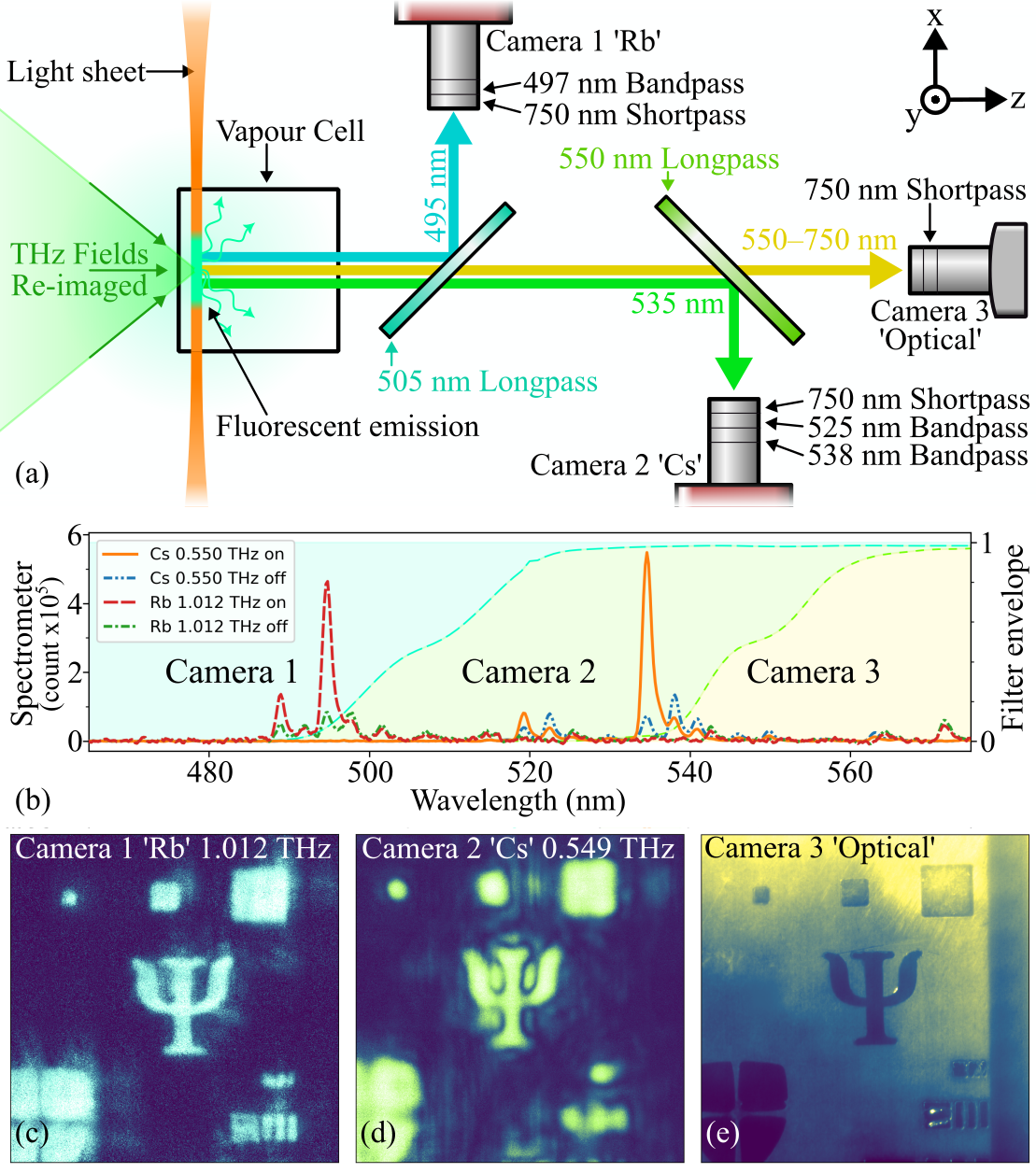}
\captionof{figure}{A schematic of the spectral separation of the fluorescence emission from atomic vapour in the science cell (a). The excited ${}^{133}$Cs and ${}^{85}$Rb atoms fluoresce at 535 nm and 495 nm light in response to the presence of 0.549\,\unit{\THz} and 1.012\,\unit{\THz} fields, respectively. 
Dichroic mirrors (dash cyan \& green lines) are used to direct wavelength-bands (coloured regions) containing the Rb and Cs fluorescence peaks towards cameras 1 \& 2, while the remaining optical light is captured by camera 3 (b). These wavelength bands are filtered further with narrow-band filters, with the resulting data captured by each camera and processed is shown in (c--e).}
\label{fig:expt_spectrum}
\labeltext{\ref{fig:expt_spectrum}\,a}{fig:expt_spectrum_sep}
\labeltext{\ref{fig:expt_spectrum}\,b}{fig:expt_spectrum_spectra}
\labeltext{\ref{fig:expt_spectrum}\,c--e}{fig:expt_spectrum_images}
\end{Figure}
\noindent they are further transferred into the ${13\textrm{D}}_{5/2}$ (\Cs{}) and ${14\textrm{D}}_{5/2}$ (\Rb{}) states, from where they decay via strong optical transitions back to the ${6\textrm{P}}_{3/2}$ and ${5\textrm{P}}_{3/2}$ states, emitting light at 535\,\unit{\nm} and 495\,\unit{\nm}, respectively. This process leads to strong emission on narrow transition lines, therefore generating optical fluorescence in the vapour that is spatially correlated with the incoming terahertz fields (see Figure~\ref{expt_setup_ef}). 

The resulting emission spectra from both atomic species is shown in Figure~\ref{fig:expt_spectrum_spectra}, where the spectra from both the lower excited states (`THz off') and upper excited states (`THz on') are shown, with the emission from the ${}^{133}$Cs predominantly in the two spikes near 535\,\unit{\nm} and from the ${}^{85}$Rb predominantly in the two spikes near 495\,\unit{\nm}. This fluorescent emission is filtered with dichroic mirrors (Figure~\ref{fig:expt_spectrum_ab}) such that the light is spectrally separated into wavelength bands  --- ${\lambda}_{\textrm{Rb}}<500\,\unit{\nm}$, $500\,\unit{\nm}<{\lambda}_{\textrm{Cs}}<550\,\unit{\nm}$, ${\lambda}_{\textrm{optical}}>550\,\unit{\nm}$, represented here by the cyan, green, and yellow arrows and regions --- and directed into optical cameras, with the 495\,\unit{\nano\metre} cyan light being reflected into the `rubidium' camera (Thorlabs CC215MU cooled sCMOS), the 535\,\unit{\nano\metre} green light reflected into the `caesium' camera (Thorlabs CC215MU cooled sCMOS), and the remaining optical yellow region showing light that enters the `optical' camera (Thorlabs CS165MU CMOS). In this way spatial separation is induced between the fluorescent response to the THz field at 0.549\,\unit{\tera\hertz}, fluorescence induced by the 1.012\,\unit{\tera\hertz} field, and optical scatter from the target sample, for isolated capture.




Each imaging arm then contains additional narrowband filters to isolate THz-induced fluorescent light from background fluorescence, scatter from IR pump lasers, and atomic decay emissions longer than 750\,\unit{\nm}. 

\labeltext{\ref{fig:expt_spectrum}\,a{\textendash}c}{fig:expt_spectrum_abc}
The resulting filtered images are shown in Figure~\ref{fig:expt_spectrum_images}, for rubidium at 495\,\unit{\nm}, for caesium at 535\,\unit{\nm}, and the optical pass-through >550\,\unit{\nm}. These are images of the test card shown in Figure~\ref{fig:expt_setup_testcard}, swept by the motorised translation stage to expand the field-of-view beyond the constraints of the 10\,\unit{\mm}-wide vapour cell, with sub-images stitched together (see Supplementary Document Section 2 for details on this processing).

\section{Image processing for materials analysis}

To highlight the benefit of simultaneous multispectral imaging, we use the system to perform materials analysis of sugar samples. The samples used here are comprised of 11\,\unit{\mm} diameter copper discs filled with the sugars glucose, maltose, lactose, ground and compressed into pellets. For comparison a similar sample of 100\,\% PTFE was also included. These samples were mounted in a 3d-printed polylactic acid (PLA) mount, affixed to a X-Y motorised translation stage for automated remote-control during the acquisition process. 

To enable an extended field of view, images were captured piecewise by moving the target samples within the imaging plane (X-Y), each with an exposure of 100\,\unit{\ms}. To allow for image processing, each camera captured two images at each target position: one image with the THz fields active, and a second image with the fields inactive (`$\textrm{THz}_{ \textrm{on}}$' and `$\textrm{THz}_{ \textrm{off}}$' for each field state, respectively), while a third camera captures an optical image of the sample. Additionally, a single pair of `brightfield' images (again with the THz fields active and inactive, `Brightfield $\textrm{THz}_{ \textrm{on}}$' and `Brightfield $\textrm{THz}_{ \textrm{off}}$') were captured with the target sample moved out of the field of view to capture variation in the THz field illumination without obstruction. These images were combined by way of the following algorithm:

\begin{equation}
    {\textrm{Image}}_{\textrm{output}} = \frac{\left( \frac{ {\textrm{THz}_{ \textrm{on}}}}{{\textrm{THz}_{ \textrm{off}}}}  - 1 \right) }{ \left( \frac{ {\textrm{Brightfield~THz}_{ \textrm{on}}}}{{\textrm{Brightfield~THz}_{ \textrm{off}}}}  - 1 \right) },
\label{eq:field_compensation}
\end{equation} 

...where the division of the $\textrm{THz}_{ \textrm{on}}$ image by the $\textrm{THz}_{ \textrm{off}}$ image compensates for variation in the light-sheet intensity both spatially over the field of view, and temporally with drifts in the pump laser power. Similarly, the denominator compensates for uneven THz field illumination in the object imaging plane. Examples of the resulting processed images can be seen in Figure~\ref{fig:results_phasemap_ab}. Here we see a clear discrimination in the ratios of transmission at the two THz frequencies between the various samples; glucose, maltose, and lactose are easily distinguished.

To highlight the spectral difference between the two images we generate a vector map using the individual Cs and Rb fluorescence fields (Figure~\ref{fig:results_phasemap_ab}) as the X and Y co-ordinates of a vector for each pixel pair in the sampled images. Wherever the pixel values in both Cs and Rb fields are similar, this generates vectors that lie along the diagonal of this vector space. Pixels that are different between the Cs and Rb fields instead generate vectors that lie in off-diagonal positions (Figure~\ref{fig:results_phasemap_c}). If we represent these vectors with a magnitude $|\textbf{R}|$ and a phase angle $\phi$, we can colour-code an output map to highlight spectral discrimination between the input Cs and Rb fields. We use a Hue-Saturation-Lightness (HSL) colour-space mapping:
\begin{equation}
    \begin{aligned}
        H &= \textrm{sign}( \phi - {\phi}_{0} ) \times |\phi - {\phi}_{0}|^{\alpha} \times A, \\
        S &= {| \phi - {\phi}_{0} |} \times B, \\
        L &= |R|,
    \end{aligned}
    \label{eq:HSLmapping}
\end{equation}
...where $\alpha$ and $A$ act as gain and gamma scaling of the the colour-wheel arc mapped to differences between transmission of the two THz fields, $B$ acts as a gain control for the colourmap saturation to limit colour to image areas with a difference in transmission, and $\phi_{0}$ an offset on the colour-wheel. These parameters have been tuned (with $A = 1.65$, $\alpha = 0.5$, $B = 7$, $\phi_{0} = 0.15$) to highlight the differences between the input samples at the imaging frequencies of 0.549\,\unit{\tera\hertz} and 1.012\,\unit{\tera\hertz}.  Spectral analysis of these samples shows that at 0.549\,\unit{\tera\hertz} glucose has the highest absorption, while maltose and lactose have similar but slightly lower absorption than glucose; and at 1.012\,\unit{\tera\hertz} the absorption of glucose rises significantly, maltose has an even greater increase in absorption than glucose, while lactose shows a significantly lower increase in absorption (Supplemental Document Section 1 contains broadband terahertz spectra of these sugar samples recorded using a Toptica Terascan photomixer system). The resulting coloured phasemap is shown in Figure~\ref{fig:results_phasemap_d}.

\begin{figure}[t]
\centering
\includegraphics[width=\linewidth]{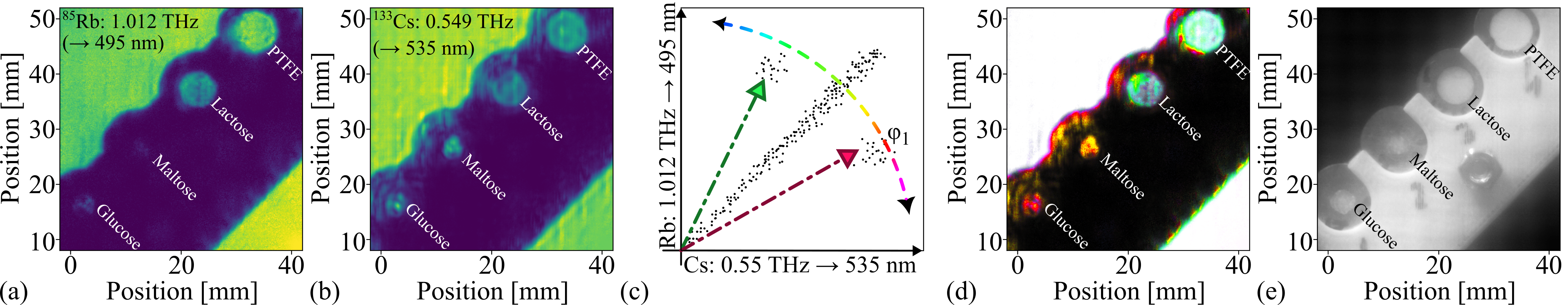}
\caption{Individual THz fields at 0.549\,\unit{\THz} (a) and 1.012\,\unit{\THz} (b), are used as X and Y co-ordinates, respectively, in a vector space (c). For each pixel pair, the resulting vector phase and magnitude is mapped to an HSL colour map. The resulting phasemap coloured by pixel phase (d), with the corresponding optical image is shown in (e).}
\label{fig:results_phasemap}
\labeltext{\ref{fig:results_phasemap}a,b}{fig:results_phasemap_ab}
\labeltext{\ref{fig:results_phasemap}c}{fig:results_phasemap_c}
\labeltext{\ref{fig:results_phasemap}d}{fig:results_phasemap_d}
\labeltext{\ref{fig:results_phasemap}e}{fig:results_phasemap_e}
\end{figure}

Image processing was performed on a laboratory computer with the Python Libraries Numpy~\cite{harris2020array}, Scipy~\cite{2020SciPy-NMeth}, and Pillow~\cite{clark2015pillow}. With modest computation power the data processing shown in Figure~\ref{fig:results_phasemap} was performed in 62\,\unit{\second}. Of this, by far the largest processing cost was in stitching individual raster-scanned images into a single ~$42\,\unit{\mm} \times 42\,\unit{\mm}$ `widefield' image for each of the two THz fields. Remaining processing required approximately 200\,\si{\ms}. When processing non-rastered data with a field-of-view limited by the vapour cell size (in this case $10\,\unit{\mm} \times 10\,\unit{\mm}$), this processing time falls to 34\,\unit{\ms}, would could be further reduced to 12\,\unit{\ms} with the use of hardware binning and cropping performed on-camera during capture. In this raster-less configuration the field of view could still be extended with the construction of a larger vapour cell~\cite{zhang202550mmtimes50}.

Simultaneous to this entire process is the `optical' field, to which only the widefield stitching algorithm has been applied in the data shown in Figure~\ref{fig:results_phasemap_e}. The atomic vapour is only non-transparent at wavelengths where an atomic transition is possible. This reduces the absorbed wavelengths to (1) the transitions which exist within the atom, and (2) to those which are adjacent to an atomic state that is populated. As a result, the atomic vapour remains largely optically transparent, and the optical camera captures light determined by the dichroic mirrors used to separate the atomic fluorescence. This does result in `background' atomic fluorescence from the lower excited ${14\textrm{P}}_{3/2}$ (Cs) and ${15\textrm{P}}_{3/2}$ (Rb) states leaking into the optical images. This could be compensated for by pulsing the optical target illumination during the 'THz off' state, such that two optical images can be compared to perform common-mode rejection of background fluorescence as has been applied to the THz fluorescence in this work. Alternatively, the IR pump lasers used to excite the atomic vapour could be pulsed to deactivate fluorescence entirely during optical field capture, at the cost of adding a third `state' to the duty cycle of image capture (in addition to 'THz field on' and 'THz field off'). 

\section{Conclusion and Outlook}

We have demonstrated simultaneous multispectral terahertz imaging using a two-atomic-species quantum sensor. The technique allows for rapid non-destructive materials analysis by comparing multiple two-point spectra in a 2D focal-plane-array format. Furthermore, we demonstrated the ability to simultaneously collect optical overlay images in the same optical pathway used by the THz imaging, due to the optically-transparent nature of the `atomic-sensor'. This could enable the addition of THz imaging to a range of optical instruments, potentially leading to the construction of new hybrid quantum sensors.

\section{Acknowledgements}

We thank Paul Dean, Alex Valavanis, Sanchit Kondawar,  Andrew Burnett and Thomas Gill of the Electronics Engineering Department of Leeds University for stimulating discussions and for preparing the sugar samples analysed in this work. We thank Cyril Bourgenot and Paul White of the Centre for Applied Instrumentation and Lee Bainbridge of the Physics Department at Durham University for aid in polishing the THz imaging lenses used in this work. We thank Malcolm Richardson of the Chemistry Department of Durham University for aid in the construction and filling of the vapour cells used in this work. We acknowledge funding from the UK Engineering and Physical Sciences Research Council grants EP/W033054/1 and EP/Z533166/1. 

\end{multicols}
\begin{multicols}{2}

\bibliography{sample}

@article{gowen2007,
  title={Hyperspectral imaging—an emerging process analytical tool for food quality and safety control},
  author={Gowen, Aoife A. and O'Donnell, Colm P. and Cullen, Patrick J. and Downey, Gerard and Frias, Jose M.},
  journal={Trends in Food Science \& Technology},
  volume={18},
  number={12},
  pages={590--598},
  year={2007},
  publisher={Elsevier}
}

@article{lu2014,
  title={Medical hyperspectral imaging: a review},
  author={Lu, Guolan and Fei, Baowei},
  journal={Journal of Biomedical Optics},
  volume={19},
  number={1},
  pages={010901},
  year={2014},
  publisher={SPIE}
}

@article{federici2005,
  title={THz imaging and sensing for security applications—explosives, weapons and drugs},
  author={Federici, John F. and Schulkin, Brian and Huang, Frank and Gary, Dale and Barat, Robert and Oliveira, Daniel and Zimdars, David},
  journal={Semiconductor Science and Technology},
  volume={20},
  number={7},
  pages={S266},
  year={2005},
  publisher={IOP Publishing}
}

@article{tonouchi2007,
  title={Cutting-edge terahertz technology},
  author={Tonouchi, Masayoshi},
  journal={Nature Photonics},
  volume={1},
  number={2},
  pages={97--105},
  year={2007},
  publisher={Nature Publishing Group}
}

@article{pickwell2006,
  title={Biomedical applications of terahertz technology},
  author={Pickwell, Emma and Wallace, Vincent P.},
  journal={Journal of Physics D: Applied Physics},
  volume={39},
  number={17},
  pages={R301},
  year={2006},
  publisher={IOP Publishing}
}

@article{chan2007,
  title={THz spectroscopy and imaging for defense and security applications},
  author={Chan, Wai Lam and Deibel, Jason and Mittleman, Daniel M.},
  journal={Reports on Progress in Physics},
  volume={70},
  number={8},
  pages={1325},
  year={2007},
  publisher={IOP Publishing}
}

@article{jepsen2011,
  title={Terahertz spectroscopy and imaging–Modern techniques and applications},
  author={Jepsen, Peter Uhd and Cooke, David G. and Koch, Martin},
  journal={Laser \& Photonics Reviews},
  volume={5},
  number={1},
  pages={124--166},
  year={2011},
  publisher={Wiley}
}

@article{kojima2017,
  title={Real-time terahertz imaging using a 320×240 microbolometer focal-plane array},
  author={Kojima, Seiji and Suzuki, Masaru and Okada, Tomoyuki and Kawase, Kodo},
  journal={Optics Express},
  volume={25},
  number={7},
  pages={8511--8519},
  year={2017},
  publisher={Optica Publishing Group}
}

@article{chan2014,
  title={A review of passive millimeter-wave imaging technology},
  author={Chan, Wai Lam and Rebeiz, Gabriel M. and Mittleman, Daniel M.},
  journal={IEEE Transactions on Microwave Theory and Techniques},
  volume={62},
  number={11},
  pages={2787--2800},
  year={2014},
  publisher={IEEE}
}

@article{wade2017,
    title={Real-time near-field terahertz imaging with atomic optical fluorescence},
    author={Wade, C. G. and \v{S}ibali{\'{c}}, N. and de Melo, N. R. and Kondo, J. M. and Adams, C. S. and Weatherill, K. J.},
    journal={Nature Photonics},
    volume={11},
    issue={1},
    pages={40},
    year={2017},
    month={Jan},
    publisher = {Nature Publishing},
    doi={10.1038/nphoton.2016.214}
}

@article{downes2020,
    title = {Full-Field Terahertz Imaging at Kilohertz Frame Rates Using Atomic Vapor},
    author = {Downes, Lucy A. and MacKellar, Andrew R. and Whiting, Daniel J. and Bourgenot, Cyril and Adams, Charles S. and Weatherill, Kevin J.},
    journal = {Phys. Rev. X},
    volume = {10},
    issue = {1},
    pages = {011027},
    numpages = {7},
    year = {2020},
    month = {Feb},
    publisher = {American Physical Society},
    doi = {10.1103/PhysRevX.10.011027}
}

@article{downes2023,
doi = {10.1088/1367-2630/acb80c},
url = {https://dx.doi.org/10.1088/1367-2630/acb80c},
year = {2023},
month = {mar},
publisher = {IOP Publishing},
volume = {25},
number = {3},
pages = {035002},
author = {Lucy A Downes and Lara Torralbo-Campo and Kevin J Weatherill},
title = {A practical guide to terahertz imaging using thermal atomic vapour},
journal = {New Journal of Physics}
}

@article{sibalic2017,
    title = {ARC: An open-source library for calculating properties of alkali Rydberg atoms},
    author = {\v{S}ibali{\'{c}}, Nikola and Pritchard, Jonathan D. and Adams, Charles S. and Weatherill, Kevin J.},
    journal = {Comp. Phys. Commun.},
    volume = {220},
    pages = {319},
    numpages = {13},
    year = {2017},
    month = {Nov},
    publisher = {Elsevier},
    doi = {10.1016/j.cpc.2017.06.015}
}

@article{chen2022,
author = {Shuying Chen and Dominic J. Reed and Andrew R. MacKellar and Lucy A. Downes and Nourah F. A. Almuhawish and Matthew J. Jamieson and Charles S. Adams and Kevin J. Weatherill},
journal = {Optica},
number = {5},
pages = {485--491},
publisher = {Optica Publishing Group},
title = {Terahertz electrometry via infrared spectroscopy of atomic vapor},
volume = {9},
month = {May},
year = {2022},
doi = {10.1364/OPTICA.456761},
}

@article{Fleming25,
author = {James P. Fleming and Lucy A. Downes and John M. Girkin and Kevin J. Weatherill},
journal = {Opt. Express},
number = {12},
pages = {26509--26516},
publisher = {Optica Publishing Group},
title = {Virtually structured illumination for terahertz super-resolution imaging},
volume = {33},
month = {Jun},
year = {2025},
url = {https://opg.optica.org/oe/abstract.cfm?URI=oe-33-12-26509},
doi = {10.1364/OE.563675},
}

@article{Downes22,
author = {Lucy A. Downes and Daniel J. Whiting and C. Stuart Adams and Kevin J. Weatherill},
journal = {Opt. Lett.},
number = {22},
pages = {6001--6004},
publisher = {Optica Publishing Group},
title = {Rapid readout of terahertz orbital angular momentum beams using atom-based imaging},
volume = {47},
month = {Nov},
year = {2022},
url = {https://opg.optica.org/ol/abstract.cfm?URI=ol-47-22-6001},
doi = {10.1364/OL.476945},
}

@ARTICLE{downes25,
  author={Downes, Lucy A. and Adams, C. Stuart and Weatherill, Kevin J.},
  journal={IEEE Transactions on Terahertz Science and Technology}, 
  title={Temporally-multiplexed Dual-frequency Terahertz Imaging at Kilohertz Frame Rates}, 
  year={2025},
  volume={},
  number={},
  pages={1-5},
  doi={10.1109/TTHZ.2025.3617783}}

@ARTICLE{Udayanga2024,
  author={Udayanga, Darsha and Serasinghe, Ashan and Dassanayake, Supun and Godaliyadda, Roshan and Herath, Vijitha and Ekanayake, Mervyn Parakrama and Malshan, Pasindu},
  journal={IEEE Transactions on Instrumentation and Measurement}, 
  title={Dual-Mode Multispectral Imaging System for Food and Agricultural Product Quality Estimation}, 
  year={2024},
  volume={73},
  number={},
  pages={1-12},
  doi={10.1109/TIM.2024.3369129}}

@Article{Lee2023,
AUTHOR = {Lee, Ki-Seung},
TITLE = {Multispectral Food Classification and Caloric Estimation Using Convolutional Neural Networks},
JOURNAL = {Foods},
VOLUME = {12},
YEAR = {2023},
NUMBER = {17},
ARTICLE-NUMBER = {3212},
URL = {https://www.mdpi.com/2304-8158/12/17/3212},
PubMedID = {37685145},
ISSN = {2304-8158},
DOI = {10.3390/foods12173212}
}

@inproceedings{Kvaterniuk2019,
author = {Serhii Kvaterniuk and Olena Kvaterniuk and Vasil Petruk and Anastasiia Mandebura and Svyatoslav Mandebura and Żaklin M. Grądz and Saule Rakhmetullina and Mukaddas Arshidinova},
title = {{Multispectral environmental monitoring of phytoplankton pigment parameters in aquatic environments}},
volume = {11176},
booktitle = {Photonics Applications in Astronomy, Communications, Industry, and High-Energy Physics Experiments 2019},
editor = {Ryszard S. Romaniuk and Maciej Linczuk},
organization = {International Society for Optics and Photonics},
publisher = {SPIE},
pages = {111762R},
year = {2019},
doi = {10.1117/12.2536809},
URL = {https://doi.org/10.1117/12.2536809}
}

@INPROCEEDINGS{Arnold2010,
  author={Arnold, Thomas and De Biasio, Martin and Fritz, Andreas and Leitner, Raimund},
  booktitle={SENSORS, 2010 IEEE}, 
  title={UAV-based multispectral environmental monitoring}, 
  year={2010},
  volume={},
  number={},
  pages={995-998},
  doi={10.1109/ICSENS.2010.5690923}}

@INPROCEEDINGS{Vetrekar2024,
  author={Vetrekar, Narayan and Ramachandra, Raghavendra and Venkatesh, Sushma and Pawar, Jyoti D. and Gad, R. S.},
  booktitle={2024 IEEE Applied Sensing Conference (APSCON)}, 
  title={Does complimentary information from multispectral imaging improve face presentation attack detection?}, 
  year={2024},
  volume={},
  number={},
  pages={1-4},
  doi={10.1109/APSCON60364.2024.10466228}}

@Article{Goessinger2024,
author={Goessinger, Elisabeth Victoria
and Dittrich, Paul-Gerald
and N{\"o}cker, Philipp
and Notni, Gunther
and Weber, Sebastian
and Cerminara, Sara
and M{\"u}hleisen, Beda
and Navarini, Alexander A.
and Maul, Lara Valeska},
title={Classification of melanocytic lesions using direct illumination multispectral imaging},
journal={Scientific Reports},
year={2024},
month={Aug},
day={16},
volume={14},
number={1},
pages={19036},
issn={2045-2322},
doi={10.1038/s41598-024-69773-x},
url={https://doi.org/10.1038/s41598-024-69773-x}
}

@article{Feiyan2023,
title = {Multispectral imaging: Review of current applications},
journal = {Survey of Ophthalmology},
volume = {68},
number = {5},
pages = {889-904},
year = {2023},
issn = {0039-6257},
doi = {https://doi.org/10.1016/j.survophthal.2023.06.004},
url = {https://www.sciencedirect.com/science/article/pii/S0039625723000851},
author = {Feiyan Ma and Mingzhen Yuan and Igor Kozak}
}

@article{Ozkan2023,
author = {Ozkan, Haydar and Aydin, Muberra and Ozcan, Osman Alpcan and Zengin, Ummuhan},
doi = {10.2478/jee-2023-0008},
url = {https://doi.org/10.2478/jee-2023-0008},
title = {A portable multispectral vein imaging system},
journal = {Journal of Electrical Engineering},
number = {1},
volume = {74},
year = {2023},
pages = {64--69}
}

@Article{VanBeek2013,
AUTHOR = {Van Beek, Jonathan and Tits, Laurent and Somers, Ben and Coppin, Pol},
TITLE = {Stem Water Potential Monitoring in Pear Orchards through WorldView-2 Multispectral Imagery},
JOURNAL = {Remote Sensing},
VOLUME = {5},
YEAR = {2013},
NUMBER = {12},
PAGES = {6647--6666},
URL = {https://www.mdpi.com/2072-4292/5/12/6647},
ISSN = {2072-4292},
DOI = {10.3390/rs5126647}
}

@Article{Stow2019,
AUTHOR = {Stow, Daniel and Nichol, Caroline J. and Wade, Tom and Assmann, Jakob J. and Simpson, Gillian and Helfter, Carole},
TITLE = {Illumination Geometry and Flying Height Influence Surface Reflectance and NDVI Derived from Multispectral UAS Imagery},
JOURNAL = {Drones},
VOLUME = {3},
YEAR = {2019},
NUMBER = {3},
ARTICLE-NUMBER = {55},
URL = {https://www.mdpi.com/2504-446X/3/3/55},
ISSN = {2504-446X},
DOI = {10.3390/drones3030055}
}

@article{Burgos-Fernandez2022,
author = {Francisco J. Burgos-Fern\'{a}ndez and Tommaso Alterini and Fernando D\'{i}az-Dout\'{o}n and Laura Gonz\'{a}lez and Carlos Mateo and Clara Mestre and Jaume Pujol and Meritxell Vilaseca},
journal = {Biomed. Opt. Express},
number = {6},
pages = {3504--3519},
publisher = {Optica Publishing Group},
title = {Reflectance evaluation of eye fundus structures with a visible and near-infrared multispectral camera},
volume = {13},
month = {Jun},
year = {2022},
url = {https://opg.optica.org/boe/abstract.cfm?URI=boe-13-6-3504},
doi = {10.1364/BOE.457412},
}

@INPROCEEDINGS{Wu2024,
  author={Wu, Shao-Hsuan and Latifi, Seyed Mostafa and Mai, Chia-Ming and Yang, Shang-Hua},
  booktitle={2024 49th International Conference on Infrared, Millimeter, and Terahertz Waves (IRMMW-THz)}, 
  title={A Hybrid Optical-Electrical Neural Network for Terahertz Computational Imaging}, 
  year={2024},
  volume={},
  number={},
  pages={1-2},
  doi={10.1109/IRMMW-THz60956.2024.10697722}}

@INPROCEEDINGS{Yehao2024,
  author={Ma, Yehao and Shen, Liran and Cao, Yuqi and Hou, Dibo and Zhang, Guangxin},
  booktitle={2024 49th International Conference on Infrared, Millimeter, and Terahertz Waves (IRMMW-THz)}, 
  title={Aerogel defect detection using terahertz waves}, 
  year={2024},
  volume={},
  number={},
  pages={1-2},
  doi={10.1109/IRMMW-THz60956.2024.10697728}}

@article{Medjadba2025,
author = {Hocine Medjadba and Hamid Zehaf and Mohamed Lazoul},
title = {{Concealed objects detection and recognition in active transmission sub-terahertz imaging scanner for automatic security screening}},
volume = {64},
journal = {Optical Engineering},
number = {7},
publisher = {SPIE},
pages = {079801},
year = {2025},
doi = {10.1117/1.OE.64.7.079801},
URL = {https://doi.org/10.1117/1.OE.64.7.079801}
}

@article{Bratu2024,
title = {Infrared to terahertz identification of chemical substances used for the production of IEDs},
journal = {Spectrochimica Acta Part A: Molecular and Biomolecular Spectroscopy},
volume = {312},
pages = {124055},
year = {2024},
issn = {1386-1425},
doi = {https://doi.org/10.1016/j.saa.2024.124055},
url = {https://www.sciencedirect.com/science/article/pii/S138614252400221X},
author = {A.M. Bratu and M. Bojan and C. Popa and M. Petrus}
}

@article{Singh2022,
title = {Dual spectroscopic detection of THz energy modes of critical chemical compounds},
journal = {Spectrochimica Acta Part A: Molecular and Biomolecular Spectroscopy},
volume = {271},
pages = {120923},
year = {2022},
issn = {1386-1425},
doi = {https://doi.org/10.1016/j.saa.2022.120923},
url = {https://www.sciencedirect.com/science/article/pii/S1386142522000713},
author = {Khushboo Singh and Diksha Garg and Aparajita Bandyopadhyay and Amartya Sengupta}
}

@article{Mathialagan2025,
author = {Meenaakumari Mathialagan and Manikandan Esakkimuthu},
title = {{Study on applicability of terahertz imaging for microplastic identification in soil sample}},
volume = {64},
journal = {Optical Engineering},
number = {6},
publisher = {SPIE},
pages = {064107},
year = {2025},
doi = {10.1117/1.OE.64.6.064107},
URL = {https://doi.org/10.1117/1.OE.64.6.064107}
}

@INPROCEEDINGS{Wang2018,
  author={Wang, Qigejian and Goay, Amus Chee Yuen and Mishra, Deepak and Atakaramians, Shaghik},
  booktitle={2024 49th International Conference on Infrared, Millimeter, and Terahertz Waves (IRMMW-THz)}, 
  title={Convolutional Neural Network Based Terahertz Imaging for Detecting Grass Seed Infestation}, 
  year={2024},
  volume={},
  number={},
  pages={1-2},
  doi={10.1109/IRMMW-THz60956.2024.10697594}}

@INPROCEEDINGS{Bertling2024,
  author={Bertling, K. and Torniainen, J. and Rakić, A.D. and Bandyopadhyay, A. and Chanda, P. and Sengupta, A. and Dean, P and Indjin, D and Li, L.H. and Davies, A.G. and Linfield, E.H.},
  booktitle={2024 49th International Conference on Infrared, Millimeter, and Terahertz Waves (IRMMW-THz)}, 
  title={Terahertz Quantum Cascade Laser Feedback Imaging for Cultural Heritage Preservation}, 
  year={2024},
  volume={},
  number={},
  pages={1-2},
  doi={10.1109/IRMMW-THz60956.2024.10697894}}

@ARTICLE{Vazquez2024,
  author={Vazquez, Danae Antunez and Pilozzi, Laura and DelRe, Eugenio and Conti, Claudio and Missori, Mauro},
  journal={IEEE Transactions on Terahertz Science and Technology}, 
  title={Terahertz Imaging Super-Resolution for Documental Heritage Diagnostics}, 
  year={2024},
  volume={14},
  number={4},
  pages={455-465},
  doi={10.1109/TTHZ.2024.3410674}}

@inproceedings{Groves2009,
author = {Roger M. Groves and Boris Pradarutti and Eleni Kouloumpi and Wolfgang Osten and Gunther Notni},
title = {{Multi-sensor evaluation of a wooden panel painting using terahertz imaging and shearography}},
volume = {7391},
booktitle = {O3A: Optics for Arts, Architecture, and Archaeology II},
editor = {Luca Pezzati and Renzo Salimbeni},
organization = {International Society for Optics and Photonics},
publisher = {SPIE},
pages = {73910E},
year = {2009},
doi = {10.1117/12.827528},
URL = {https://doi.org/10.1117/12.827528}
}

@article{Yan2022,
title = {THz medical imaging: from in vitro to in vivo},
journal = {Trends in Biotechnology},
volume = {40},
number = {7},
pages = {816-830},
year = {2022},
issn = {0167-7799},
doi = {https://doi.org/10.1016/j.tibtech.2021.12.002},
url = {https://www.sciencedirect.com/science/article/pii/S016777992100295X},
author = {Zhiyao Yan and Li-Guo Zhu and Kun Meng and Wanxia Huang and Qiwu Shi}
}

@ARTICLE{Taylor2011,
  author={Taylor, Zachary D. and Singh, Rahul S. and Bennett, David B. and Tewari, Priyamvada and Kealey, Colin P. and Bajwa, Neha and Culjat, Martin O. and Stojadinovic, Alexander and Lee, Hua and Hubschman, Jean-Pierre and Brown, Elliott R. and Grundfest, Warren S.},
  journal={IEEE Transactions on Terahertz Science and Technology}, 
  title={THz Medical Imaging: in vivo Hydration Sensing}, 
  year={2011},
  volume={1},
  number={1},
  pages={201-219},
  doi={10.1109/TTHZ.2011.2159551}}

@inproceedings{Sung2012,
author = {Shijun Sung and Neha Bajwa and Nuhba Fokwa and Priyamvada Tewari and Rahul Singh and Martin Culjat and Bryan Nowroozi and Warren Grundfest and Zachary Taylor},
title = {{Fast-scanning THz medical imaging system for clinical application}},
volume = {8496},
booktitle = {Terahertz Emitters, Receivers, and Applications III},
editor = {Manijeh Razeghi and Alexei N. Baranov and Henry O. Everitt and John M. Zavada and Tariq Manzur},
organization = {International Society for Optics and Photonics},
publisher = {SPIE},
pages = {84960S},
year = {2012},
doi = {10.1117/12.946013},
URL = {https://doi.org/10.1117/12.946013}
}

@article{Kawase2003,
author = {Kodo Kawase and Yuichi Ogawa and Yuuki Watanabe and Hiroyuki Inoue},
journal = {Opt. Express},
number = {20},
pages = {2549--2554},
publisher = {Optica Publishing Group},
title = {Non-destructive terahertz imaging of illicit drugs using spectral fingerprints},
volume = {11},
month = {Oct},
year = {2003},
url = {https://opg.optica.org/oe/abstract.cfm?URI=oe-11-20-2549},
doi = {10.1364/OE.11.002549},
}

@ARTICLE{Eliet2022,
  author={Eliet, Sophie and Cuisset, Arnaud and Hindle, Francis and Lampin, Jean-François and Peretti, Romain},
  journal={IEEE Transactions on Terahertz Science and Technology}, 
  title={Broadband Super-Resolution Terahertz Time-Domain Spectroscopy Applied to Gas Analysis}, 
  year={2022},
  volume={12},
  number={1},
  pages={75-80},
  doi={10.1109/TTHZ.2021.3120029}}

@article{Nikolaev2020,
doi = {10.1088/1742-6596/1461/1/012118},
url = {https://dx.doi.org/10.1088/1742-6596/1461/1/012118},
year = {2020},
month = {mar},
publisher = {IOP Publishing},
volume = {1461},
number = {1},
pages = {012118},
author = {Nikolaev, N A and Rybak, A A and Kuznetsov, S A},
title = {Application of metasurface-based low-pass filters for improving THz-TDS characteristics},
journal = {Journal of Physics: Conference Series}
}

@article{Li_2025,
doi = {10.1088/1361-6463/ad9c8e},
url = {https://doi.org/10.1088/1361-6463/ad9c8e},
year = {2024},
month = {dec},
publisher = {IOP Publishing},
volume = {58},
number = {8},
pages = {085109},
author = {Li, Xianzhe and Li, Tao and Wan, Jun and Zhang, Bin and Huang, Qirong and Yang, Xinyu and Feng, Lie and Zhang, Kaiqing and Huang, Wei and Deng, Haixiao},
title = {Dual-cameras terahertz imaging with multi-kilohertz frame rates and high sensitivity via Rydberg-atom vapor},
journal = {Journal of Physics D: Applied Physics}
}

@misc{zhang202550mmtimes50,
      title={50 mm $\times$ 50 mm Cesium Atomic Vapor Cell for Terahertz Imaging: Implementation and Application}, 
      author={Bin Zhang and Jun Wan and Tao Li and Xian-Zhe Li and Yu Wu and Qi-Rong Huang and Xin-Yu Yang and Wei Huang and Kai-Qing Zhang and Hai-Xiao Deng},
      year={2025},
      eprint={2509.22098},
      archivePrefix={arXiv},
      primaryClass={physics.optics},
      url={https://arxiv.org/abs/2509.22098}, 
}

@misc{krokosz2025electricfieldmetrologyterahertzfrequency,
      title={Electric-field metrology of a terahertz frequency comb using Rydberg atoms}, 
      author={Wiktor Krokosz and Jan Nowosielski and Bartosz Kasza and Sebastian Borówka and Mateusz Mazelanik and Wojciech Wasilewski and Michał Parniak},
      year={2025},
      eprint={2508.20698},
      archivePrefix={arXiv},
      primaryClass={physics.optics},
      url={https://arxiv.org/abs/2508.20698}, 
}

@book{Foulkes-PhysChem2013,
author={Foulkes, Frank R.},
title={Physical Chemistry for Engineering and Applied Sciences},
year={2013},
edition={1},
publisher={CRC Press},
address={Milton},
isbn={9781466518469},
doi={10.1201/b12732},
url={https://doi.org/10.1201/b12732}
}

@Article{harris2020array,
 title         = {Array programming with {NumPy}},
 author        = {Charles R. Harris and K. Jarrod Millman and St{\'{e}}fan J.
                 van der Walt and Ralf Gommers and Pauli Virtanen and David
                 Cournapeau and Eric Wieser and Julian Taylor and Sebastian
                 Berg and Nathaniel J. Smith and Robert Kern and Matti Picus
                 and Stephan Hoyer and Marten H. van Kerkwijk and Matthew
                 Brett and Allan Haldane and Jaime Fern{\'{a}}ndez del
                 R{\'{i}}o and Mark Wiebe and Pearu Peterson and Pierre
                 G{\'{e}}rard-Marchant and Kevin Sheppard and Tyler Reddy and
                 Warren Weckesser and Hameer Abbasi and Christoph Gohlke and
                 Travis E. Oliphant},
 year          = {2020},
 month         = sep,
 journal       = {Nature},
 volume        = {585},
 number        = {7825},
 pages         = {357--362},
 doi           = {10.1038/s41586-020-2649-2},
 publisher     = {Springer Science and Business Media {LLC}},
 url           = {https://doi.org/10.1038/s41586-020-2649-2}
}

@ARTICLE{2020SciPy-NMeth,
  author  = {Virtanen, Pauli and Gommers, Ralf and Oliphant, Travis E. and
            Haberland, Matt and Reddy, Tyler and Cournapeau, David and
            Burovski, Evgeni and Peterson, Pearu and Weckesser, Warren and
            Bright, Jonathan and {van der Walt}, St{\'e}fan J. and
            Brett, Matthew and Wilson, Joshua and Millman, K. Jarrod and
            Mayorov, Nikolay and Nelson, Andrew R. J. and Jones, Eric and
            Kern, Robert and Larson, Eric and Carey, C J and
            Polat, {\.I}lhan and Feng, Yu and Moore, Eric W. and
            {VanderPlas}, Jake and Laxalde, Denis and Perktold, Josef and
            Cimrman, Robert and Henriksen, Ian and Quintero, E. A. and
            Harris, Charles R. and Archibald, Anne M. and
            Ribeiro, Ant{\^o}nio H. and Pedregosa, Fabian and
            {van Mulbregt}, Paul and {SciPy 1.0 Contributors}},
  title   = {{{SciPy} 1.0: Fundamental Algorithms for Scientific
            Computing in Python}},
  journal = {Nature Methods},
  year    = {2020},
  volume  = {17},
  pages   = {261--272},
  adsurl  = {https://rdcu.be/b08Wh},
  doi     = {10.1038/s41592-019-0686-2},
}

@misc{clark2015pillow,
  title={Pillow (PIL Fork) Documentation},
  author={Clark, Alex},
  year={2015},
  publisher={readthedocs},
 url={https://buildmedia.readthedocs.org/media/pdf/pillow/latest/pillow.pdf}
}

\bibliographyfullrefs{sample}
\end{multicols}

\ifthenelse{\equal{\journalref}{aop}}{%
\section*{Author Biographies}
\begingroup
\setlength\intextsep{0pt}
\begin{minipage}[t][6.3cm][t]{1.0\textwidth} 
  \begin{wrapfigure}{L}{0.25\textwidth}
    \includegraphics[width=0.25\textwidth]{john_smith.eps}
  \end{wrapfigure}
  \noindent
  {\bfseries Andrew MacKellar} received his PhD (Physics) in 2017 from The University of Strathclyde. His research interests include quantum sensors.
\end{minipage}
\begin{minipage}{1.0\textwidth}
  \begin{wrapfigure}{L}{0.25\textwidth}
    \includegraphics[width=0.25\textwidth]{alice_smith.eps}
  \end{wrapfigure}
  \noindent
  {\bfseries Alice Smith} also received her BSc (Mathematics) in 2000 from The University of Maryland. Her research interests also include lasers and optics.
\end{minipage}
\endgroup
}{}

\end{document}